\documentclass[aps,prb,superscriptaddress,twocolumn]{revtex4-1}
\usepackage{feynmp-auto}
\usepackage{graphicx,import} % Required for including pictures

\usepackage{float} % Allows putting an [H] in \begin{figure} to specify the exact location of the figure
\usepackage{wrapfig} % Allows in-line images such as the example fish picture
\usepackage{amsfonts}
\usepackage{lipsum} % Used for inserting dummy 'Lorem ipsum' text into the template

\usepackage{xr} %cross reference to other document
\usepackage{braket}
\usepackage{amsmath}
\usepackage{color}
\usepackage[pdfpagelabels]{hyperref}
\usepackage{epstopdf}
\usepackage{gensymb} %degree, for example 90{\degree}
\usepackage{xr} %refer external document
\DeclareMathOperator{\tr}{tr}
\DeclareMathOperator{\erf}{erf}
\DeclareMathOperator{\sgn}{sgn}
\newcommand{\RN}[1]{%
  \textup{\uppercase\expandafter{\romannumeral#1}}%
}

\begin{document}
\allowdisplaybreaks
%\preprint{}

\title{Anderson orthogonality catastrophe in $2+1$-D topological systems}


\author{Jiahua Gu}
%\thanks{}
\affiliation{Department of Physics, University of Michigan, Ann Arbor, Michigan 48109, USA}


\date{\today}

\begin{abstract}
In the thermodynamic limit, a many-body ground state has zero overlap with another state which is a slightly perturbed state of the original one, known as the Anderson orthogonality catastrophe (AOC). The amplitude of the overlap for two generic ground states typically exhibits exponential or power-law decay as the system size increases to infinity, depending on whether the bulk is insulating or conducting. In this paper, we show that for generic $(2+1)$-D topological systems at fixed points, there exists a universal topological response term in the scaling of the ground-state overlap. For Laughlin wave functions, in particular, we also find a leading term decaying faster than exponential, which is beyond AOC. Such finite-size scaling behaviors could be utilized to theoretically detect the gapless edge modes, distinguish the topology of quantum states or serve as a signature for topological phase transitions.
\end{abstract}

\pacs{}
%\keywords{}

\maketitle
\section{Introduction}
Ground states of condensed matter systems encode the information of quantum phases. Topological insulators, for example, are defined through the calculation of certain topological invariants \cite{Haldane1988model, Kane2005quantum, Kane2005z2, Fu2006time, Fu2007topological1, Fu2007topological2, Bernevig2006quantum} based on ground states. There has been many efforts trying to understand such information in the ground-state wave functions. In particular, the ground-state overlaps have been utilized to investigate geometric entanglement \cite{Wei2003geometric,Blasone2008hierarchies,Orus2014geometric} and (topological) quantum phase transitions \cite{Gu2016adiabatic, Gu2010fidelity,Varney2011topological,Varney2010interaction,konig2016universal,Zanardi2006ground}. Among those efforts, our previous work \cite{Gu2016adiabatic} has proven that two insulators lie in the same topological phase if their single-particle ground-state overlap does not vanish in the first Brillouin zone.

However, many-body ground-state overlaps in the thermodynamic limit are always 0 due to Anderson orthogonality catastrophe \cite{Anderson1967infrared}. At first glance, it seems no information could be extracted from the wave-function overlap. But we learned in first-year calculus that functions could approach 0 in different order under certain limit condition. Thus, it is expected that more information could be generated from the finite-size scaling of overlaps. For the impurity problem, Anderson first showed a power law decay for the overlap,
\begin{equation}
\braket{\Psi|\Psi^\prime} \sim N^{-\epsilon},
\end{equation}
where $N$ is the number of electrons and $\epsilon>0$. For two generic wave functions, if the overlap on each lattice site differs by a finite amount then one would naively expect the many-body wave-function overlap exhibit an exponential decay.

In this paper, we focus on the study of $(2+1)$-D topological states which includes both symmetry-protected topological (SPT) states and intrinsic topological states. SPT systems are adiabatically connected with trivial product states under local unitary transformation \cite{Chen2013symmetry, Gu2014symmetry}. However, there is no such a smooth path connecting nontrivial SPT states and product states if the symmetries are preserved. Such systems in $(2+1)$-D must have gapless surface/edge modes \cite{Chen2013symmetry, Gu2014symmetry}. In higher dimensions, these boundary modes could also be gapped. But the ground state on the boundary must be degenerate if the essential symmetries are spontaneously broken there \cite{Chen2013symmetry, Gu2014symmetry}. In the following, we will exploit the ideal ground-state wave function for $(2+1)$-D fixed-point SPT systems (in renormalization-group sense) to prove the existence of a topological response term in ground-state overlaps. Then we will verify it with the example of $\mathbb Z_2$-protected Ising paramagnetic topological systems \cite{Levin2012braiding}. We show that the overlap of generic fixed-point SPT states has a universal sub-leading terms depending on the Euler characteristic of the manifold the systems live on. Such a topological response term is an analogue of the corrections of entanglement entropy and free energy in $(1+1)$-D critical systems. Indeed, we find the coefficient of the topological response term is related to the central charge of an underlying CFT. Similar behavior is seen in systems with intrinsic topological order. More surprisingly, we find an overlap decaying faster than exponential in the case of fractional quantum Hall (FQH) systems. We will end with some open questions.

\section{Physical Intuition \label{sec: physical}}
In this section, we will provide an intuitive argument that will serve as a basic physical picture for the rigorous calculations in later sections.

The imaginary time evolution of an arbitrary state could be written as
\begin{equation}
\ket{\text{final}} = \mathcal T\left(e^{-\int_0^\beta H(\tau) d\tau}\right) \ket{\text{initial}}
\label{eq: evolution}
\end{equation}
up to a normalization factor. Here $\mathcal T$ is the time-ordering operator, the initial state is an arbitrary state at time 0 and it evolves within time $\beta$ to the final state. Denote the eigenvalues and eigenstates of such a system as $E_i$ and $\ket{i}$ respectively. $\ket{i}$'s form a complete set of basis vectors in the wave-function space. So any initial state could be written as a linear combination of $\ket{i}$'s,
\begin{equation}
\ket{\text{initial}} = \sum_i \alpha_i \ket{i},
\end{equation}
where $\alpha_i$'s are finite constants. Insert this equation to Eq. (\ref{eq: evolution}), we get
\begin{equation}
\ket{\text{final}} = \sum_i \alpha_i e^{-\int_0^\beta E_i d\tau}\ket{i}
\end{equation}
It becomes obvious that in the infinite-time/zero-temperature limit $\beta \rightarrow \infty$, coefficients of all excited states are suppressed exponentially to 0. In other words, if the initial state contains some component from the ground states, then the final state must be a linear combination of degenerate ground states under long-time evolution. In the case the ground state is non-degenerate, we have reached the unique ground state in this calculation. This is consistent with the fact that when the temperature is 0, quantum systems reside in their ground states.

\begin{figure}[h]
\centering
\includegraphics[scale=0.45]{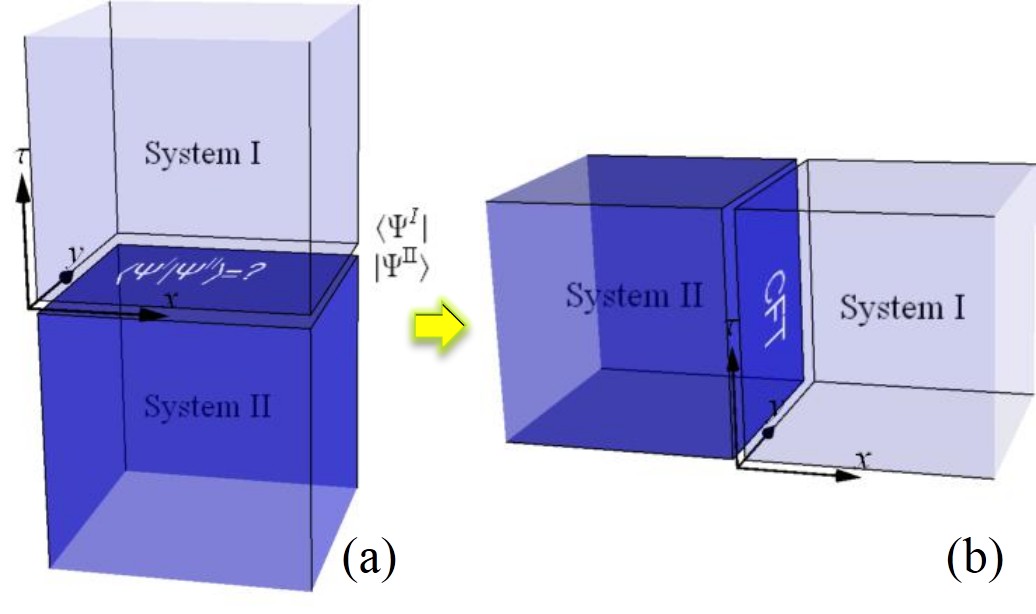}
\caption{(a) $\ket{\Psi^{\RN{1}}},\ket{\Psi^{\RN{2}}}$ are obtained from the infinite-time evolution of arbitrary states from $+\infty$ and $-\infty$ under the path integral of system $\RN{1}$ and $\RN{2}$ respectively. The overlap describes the interface $\tau=0$. (b) Rotate the combined system by $90{\degree}$. The interface becomes $x=0$ which is a CFT.}
\label{fig: physical_picture}
\end{figure}

Once the ground states are obtained, we could calculate the ground-state overlaps. Similar to the argument by You et al. \cite{You2014wave}, here we only give an intuitive argument. In the following sections, we will prove it through rigorous formulation for a large variety of fixed-point topological systems. Consider two (almost) arbitrary states evolving under the Hamiltonian of systems $\RN{1}$ and $\RN{2}$ from $\tau=+\infty$ and $\tau=-\infty$ to $\tau=0$ respectively. Since the components of excited states in the two initial states are exponentially suppressed during the evolution, the final states at $\tau=0^+$ and $\tau=0^-$ are the ground states $\bra{\Psi^{\RN{1}}}, \ket{\Psi^{\RN{2}}}$ of system $\RN{1}$ and system $\RN{2}$ respectively (Fig. \ref{fig: physical_picture}(a)). So the overlap $\braket{\Psi^{\RN{1}}|\Psi^{\RN{2}}}$ describes the theory on the interface $\tau=0$. Rotating such a system by $90{\degree}$, the imaginary-time interface at $\tau=0$ becomes a spatial interface at $x=0$ (Fig. \ref{fig: physical_picture}(b)). So as long as the interface at $x=0$ is gapless (which is true for any $(2+1)$-D SPT systems) and described by some CFT, the corresponding wave function overlap would be related to a critical theory.

Indeed, we will show later case by case that the overlaps can be interpreted as partition function $Z$ of certain CFT. For a generic CFT, Cardy \cite{Cardy1988finite} shows that the free energy scales as
\begin{equation}
F=-\ln(Z)=\alpha N+\beta \sqrt N -\frac{\chi c}{12} \ln(N)+O(1).
\label{eq: CFT_scaling}
\end{equation}
In this expression, the $N$ term comes from the contribution of bulk, the $\sqrt N$ term from the boundary contribution. The coefficient of $\ln(N)$ correction depends on the Euler characteristic $\chi$ of the manifold $M$ where the system lives on and the central charge $c$ of the underlying CFT. Following this key expression for scaling, we conclude that $\ln(\braket{\Psi^{\RN{1}}|\Psi^{\RN{2}}})$ contains the topological response term $\frac{\chi c}{12} \ln(N)$ as we expected.

\section{Bosonic SPT}
In this section, we will give a more rigorous derivation for $(2+1)$-D fixed-point bosonic SPT systems. It was shown that a large number of $(2+1)$-D SPT states could be described by \textit{continuous} nonlinear $\sigma$ models with topological $\theta$ terms \cite{Bi2015Classification}. For example, in spin systems with $SO(3)$ rotational symmetry, the action is given by
\begin{equation}
\begin{aligned}
S = & \int d\tau d^2 x \left( \frac{1}{2\rho} \tr(\partial_\mu g^\dagger \partial_\mu g) \right.\\
& \left. + i \frac{\theta}{2\pi^2} \frac{\epsilon^{\mu\nu\lambda}}{6}\frac{1}{8}\tr[(g^{-1}\partial_\mu g)(g^{-1}\partial_\nu g) (g^{-1}\partial_\lambda g)]\vphantom{\frac{1}{2\rho}}\right)
\end{aligned}
\end{equation}
where $g(\boldsymbol x, t)$ is a group element in $SO(3)$ and $\theta = 2\pi k$ with $k \in \mathbb Z$. A more complete classification of bosonic SPT can be achieved through group cohomology \cite{Chen2012symmetry, Chen2013symmetry}, which could be loosely considered as a discrete version for the topological field theories mentioned above. In the group cohomology construction, the partition function for fixed-point systems are required to be 1. This will be one of the essential requirements when we work with the discrete space-time path integrals in the following.

\subsection{Warm-up: Ground States of $(1+1)$-D Fixed Points}
\begin{figure}[h]
\centering
\includegraphics[scale=0.35]{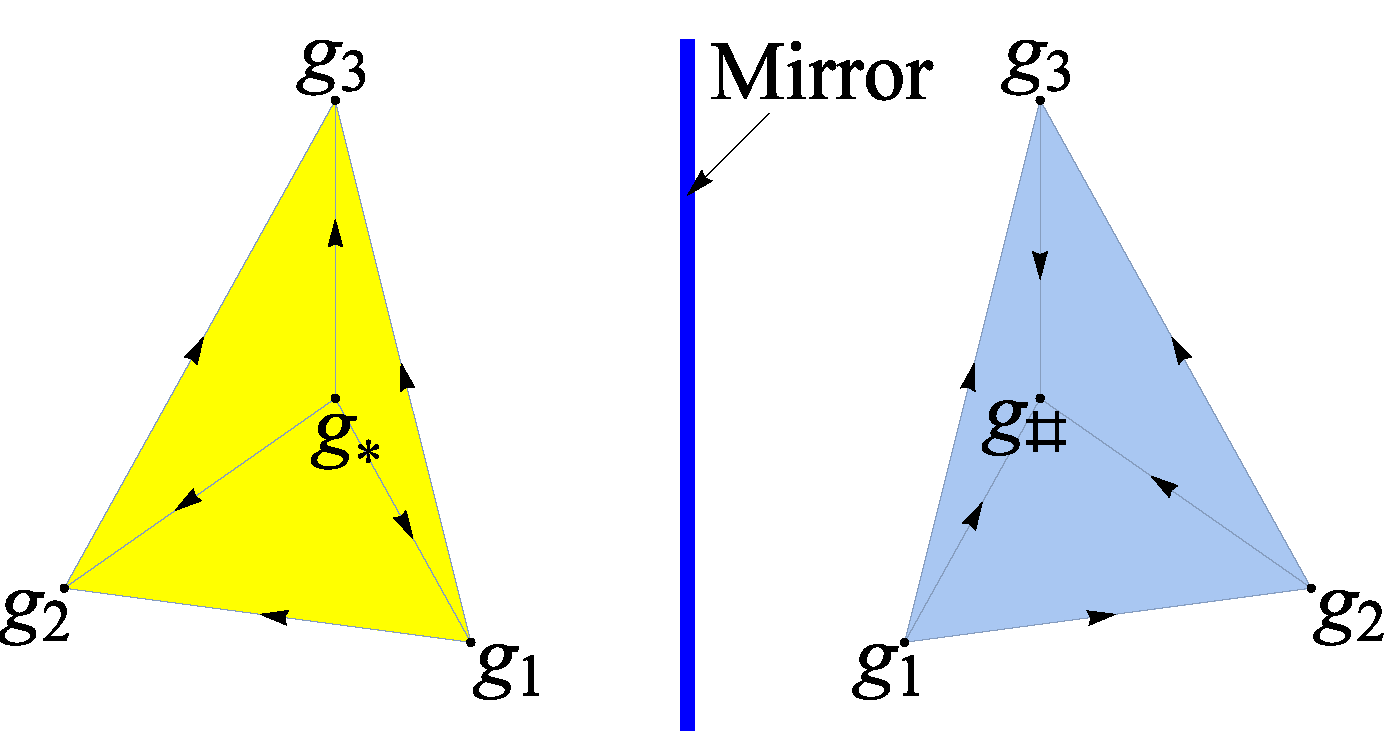}
\caption{For simplicity, we only illustrate the graphic representation of the fixed point wave function in $(1+1)$-D where $M$ is topologically a circle instead of a sphere in the main text. (Left) $\Psi(\{g_i\in M\})=|G|^{-5/2} \sum_{g_*\in G} \nu_2^{-1}(g_*,g_1,g_2) \nu_2^{-1}(g_*,g_2,g_3) \nu_2(g_*,g_1,g_3)$. The time evolution of an arbitrary state from the point $*$ to the boundary $\partial \Sigma=M$ would produce the ground state $\ket{\Psi}$ of the corresponding theory. (Right) $\Psi^\dagger(\{g_i\in M\})= |G|^{-5/2} \sum_{g_*\in G} \nu_2(g_1,g_2,g_\#) \nu_2(g_2,g_3,g_\#) \nu_2^{-1}(g_1,g_3,g_\#)$. The Hermitian conjugate $\bra{\Psi}$ can be obtained through a mirror reflection of the state $\ket{\Psi}$ except the orientation of each simplex is reversed. And such a state could be considered as some arbitrary wave function evolves from the point $\#$ backward in time to the boundary $M$.}
\label{fig: cohomology_wave_function}
\end{figure}

As we have seen in Section \ref{sec: physical}, theories in continuous space-time can find their ground-state wave function by infinite-time evolution of (almost) arbitrary states. Similarly, after we discretize the space-time, the fixed-point ground states of different SPT phases could be represented through discretized version of space-time evolution of states. For the sake of simplicity, we give an example of a $(1+1)$-D bosonic SPT sytem with onsite symmetry $G$ (see Fig. \ref{fig: cohomology_wave_function}). In the discrete space-time or triangulation of space-time, group cohomology theory assigns one group element $g\in G$ to each vertex and a phase factor $\nu$ (dubbed cocycle) to the simplex (a triangle in $(1+1)$-D and a tetrahedron in $(2+1)$-D).

Intuitively, the fixed-point ground state of such a system is given by the time evolution of an arbitrary state from vertex $*$ to the boundary $(1 2 3)$. Mathematically, the combinations of $\{g_1, g_2, g_3\}$ on the boundary forms an orthogonal set of basis vectors in the wave-function space. The ground-state amplitude over the basis vector $\ket{\{g_1, g_2, g_3\}}$ is given by
\begin{equation}
\begin{aligned}
&\Psi(\{g_1, g_2, g_3\})\\
=&\frac{1}{|G|}\sum_{g_*\in G} \nu_2^{-1}(g_*,g_1,g_2) \nu_2^{-1}(g_*,g_2,g_3) \nu_2(g_*,g_1,g_3),
\end{aligned}
\end{equation}
where $|G|$ is the order of group $G$, $\nu_2(g_i, g_j, g_k)$ is the 2-cocycle associated with triangle $(i j k)$ and its exponent $\pm 1$ is determined through the orientation of the triangle $(i j k)$. Each $\nu$ could be considered as the discrete version of the action amplitude $e^{-S}$ where $S$ is the fixed-point action of the topological system. The Hermitian conjugate of this state is represented as the mirror image of the triangulation on the left panel in Fig. \ref{fig: cohomology_wave_function} except the time-evolution arrows reversed. Its wave-function amplitude over the basis $\ket{\{g_1, g_2, g_3\}}$ is
\begin{equation}
\begin{aligned}
&\Psi^\dagger(\{g_1, g_2, g_3\})\\
=&\frac{1}{|G|}\sum_{g_\#\in G} \nu_2(g_1,g_2,g_\#) \nu_2(g_2,g_3,g_\#) \nu_2^{-1}(g_1,g_3,g_\#).
\end{aligned}
\end{equation}
Pictorially, the overlap $\braket{\Psi|\Psi} = \tr (\Psi \Psi^\dagger)$ is the gluing of two $(1+1)$-D manifolds \cite{Gu2014symmetry} sharing the same boundary $(1 2 3)$. This is mathematically represented as
\begin{equation}
\begin{aligned}
\braket{\Psi|\Psi} =\frac{1}{|G|^3}\sum_{g1, g2, g3}\Psi(\{g_1, g_2, g_3\}) \Psi^\dagger(\{g_1, g_2, g_3\})
\end{aligned}
\end{equation}
where the third power of $|G|$ is due to the fact that there are in total 3 vertices on the gluing boundary. For this specific $(1+1)$-D case, the resulting manifold is a sphere with the vertex $g_*$ representing negative infinite time and $g_{\#}$ representing positive infinite time. Plugging in the expressions of $\Psi$ and $\Psi^\dagger$, we immediately realize that  $\braket{\Psi|\Psi}$ is exactly the discrete version of a path integral. Since the path integral of fixed-point topological field theories over closed manifolds \cite{Chen2013symmetry} are required to be 1,  we obtain $\braket{\Psi|\Psi} = 1$, consistent with the normality of quantum states. In fact, this result could also be trivially derived from the more basic branching rules \cite{Chen2013symmetry}, which we omit here.

\subsection{Ground-state Overlaps of $(2+1)$-D Fixed Points}
\begin{figure}[h]
\centering
\includegraphics[scale=0.2]{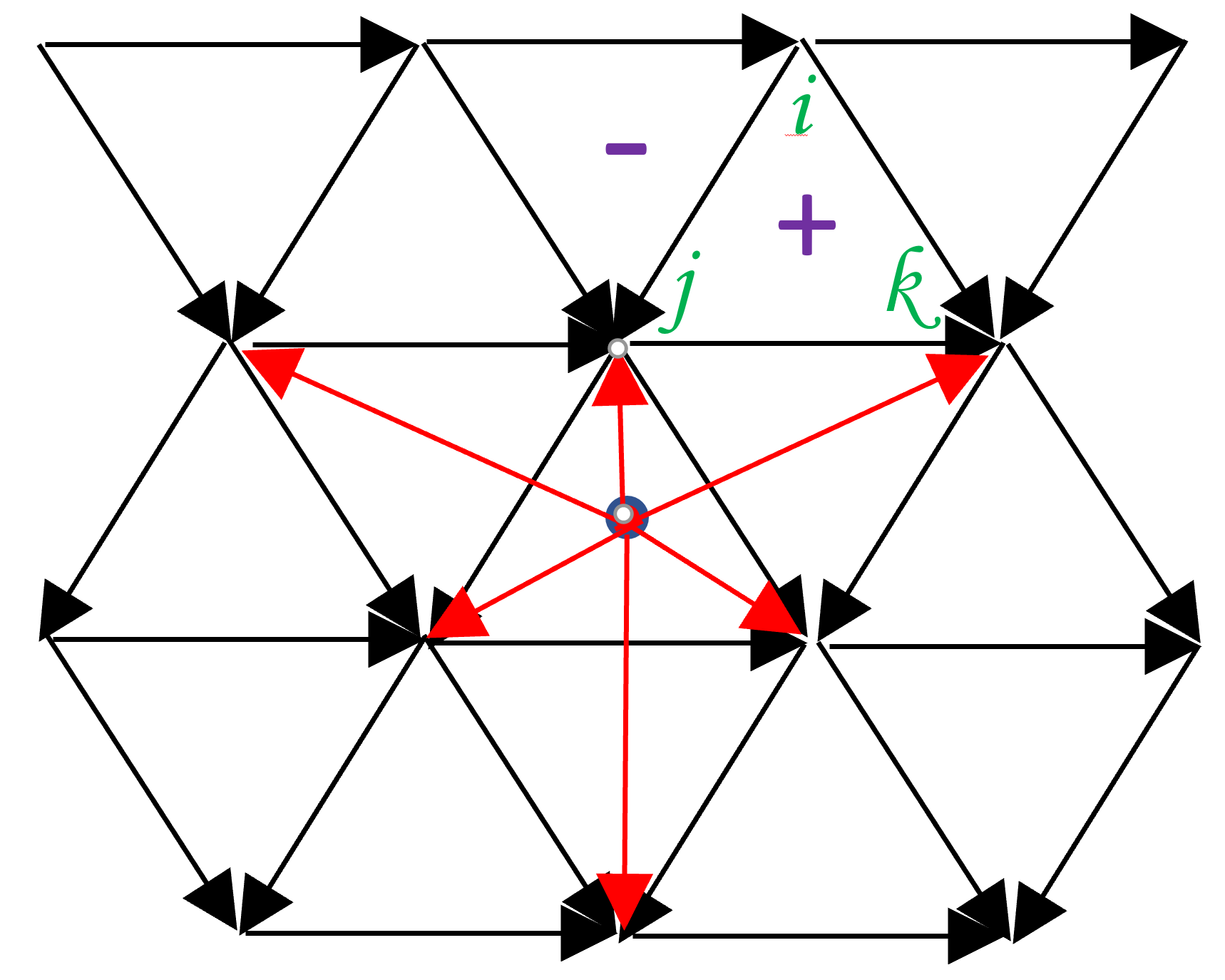}
\caption{One part of a triangulation of $(2+1)$-D manifold. The red arrows point from a point inside of $\Sigma$ to points on the boundary $M$ and the black arrows form a triangulation of $M$. Together they form a triangulation of the $(2+1)$-D manifold $\Sigma$. Associated with vertices are group elements $g$. Triangle $[ijk]$ has positive orientation while its adjacent triangles have negative orientation. The orientation of other triangles could be derived easily from this convention.}
\label{fig: triangulation}
\end{figure}

Now consider a bosonic SPT system with on-site symmetry $G$ defined on 2-dimensional closed manifold $M$. Including the time direction, the ground state would be defined on a $(2+1)$-D manifold $\Sigma$ with boundary $\partial \Sigma = M$. Pick an arbitrary point $*$ inside $\Sigma$ and connect it with all the vertices on $M$. In this way we have built a triangulation of the $(2+1)$-D manifold (see Fig. \ref{fig: triangulation}). One example for such a manifold $\Sigma$ is a solid 2-sphere and its boundary $M$ is a hollow 2-sphere. But it could be much more general. The ground-state wave function could be considered as the time evolution of an arbitrary state from any point $*$ inside of $\Sigma$ to its boundary $M$. Explicitly, the fixed-point wave-function amplitude is given by
\begin{equation}
\begin{aligned}
\Psi(\{g_i\}_M)=\frac{1}{|G|}\sum_{g_*\in G}\prod_{(*ijk)\in \Sigma} \nu_3^{s_{*ijk}}(g_*, g_i, g_j, g_k).
\end{aligned}
\end{equation}
Here $\nu_3^{s_{*ijk}}(g_*, g_i, g_j, g_k)$ corresponds to the action amplitude $e^{-\int_{(*ijk)}d^2x d\tau \mathcal L[g(\boldsymbol x,\tau)]}$ on a single simplex $(*ijk)$ and $s_{*ijk}=\pm 1$ depends on the orientation of the simplex. The summation $|G|^{-1} \sum_{g_*\in G}$ is understood as $\int dg_i$ over the group manifold if $G$ is a continuous group. We also denoted $\Sigma$ as the discrete $(2+1)$-D complex whose boundary is $M$ and inside of $\Sigma$ there is only one more vertex $*$. In order for the action $S$ to exhibit a quantized topological $\theta$-term, one essential constraint on the three-cocycle $\nu_3(g_i,g_j,g_k,g_l)\in \mathcal H^3[G,U_T(1)]$ is that on closed $(2+1)$-D manifolds,
\begin{equation}
e^{-S(\{g_i\})}=\prod_{(ijkl)} \nu_3^{s_{ijkl}}(g_i, g_j, g_k, g_l)=1.
\label{eq: cocycle_condition}
\end{equation}

The above wave function can be considered as a state on $M$ evolved from $*$. All excited states are exponentially suppressed. So such a wave function is the ground state. And its Hermitian conjugate is the state defined on the mirror reflection $\widetilde \Sigma$ (similar to the $(1+1)$-D case in Fig. \ref{fig: cohomology_wave_function}). Therefore calculating the inner product of these two states is equivalent to gluing $\Sigma$ and $\widetilde \Sigma$ together along $M$ \cite{Gu2014symmetry}. Similarly, the overlap of two distinct SPT states $\ket{\Psi^{\RN{1}}},\ket{\Psi^{\RN{2}}}$ is a gluing with mismatches of the 3-cocycle $\nu_3$,
\begin{equation}
\begin{aligned}
&\braket{\Psi_M^{\RN{1}}|\Psi_M^{\RN{2}}}=\frac{1}{|G|^{N+2}}\sum_{\{\text{all }g\text{ on }\widetilde {\Sigma}\cup \Sigma\}}\\
&\prod_{(*ijk)}(\nu_3^{\RN{1}})^{-s_{*ijk}}(g_*, g_i, g_j, g_k) \prod_{(ijk\#)}(\nu_3^{\RN{2}})^{s_{ijk \#}}(g_i,g_j,g_k,g_\#)
\end{aligned}
\label{eq: bosonic_SPT}
\end{equation}
where $N$ is the number of vertices on $M$ and thus $N+2$ is the total number of vertices including $*$ and $\#$. $\widetilde \Sigma \cup \Sigma$ becomes a $(2+1)$-D closed manifold after the gluing and the orientation has been taken care of by the sign-flip of $s_{*ijk}$. Those 3-cocycles satisfying Eq. (\ref{eq: cocycle_condition}) cancel out. Finally one side of the mirror is occupied by trivial cocycles $\nu_3(g_*, g_i, g_j, g_k)=1$ and the other side is occupied by the difference of the topological $\theta$-terms for system $\RN{1}$ and $\RN{2}$. On the boundary $M$ the discrete version of the topological $\theta$-term reduces to the Wess-Zumino-Witten (WZW) theory \cite{Chen2011two, Liu2013Symmetry, Witten1983Global, Witten1984Non}. So the final expression for the overlap calculation is identical to a partition function $Z$ of a conformal field theory (CFT) described by WZW action. 

As mentioned above, the scaling behavior of generic $(1+1)$-D CFT is known to follow Eq. (\ref{eq: CFT_scaling}). In our case the contribution from the $\sqrt N$ boundary term is 0 since $M$ is a closed manifold without boundary. Therefore the finite-size scaling of the overlap follows
\begin{equation}
\ln \braket{\Psi^{\RN{1}}|\Psi^{\RN{2}}}= -\alpha N+\frac{\chi c}{12}\ln N+O(1).
\label{eq: BSPT_scaling}
\end{equation}
The ground states are derived for generic $(2+1)$-D fixed-point bosonic SPT systems. So this result is also generic in the same realm. In $(2+1)$-D, SPT systems are guaranteed to have gapless modes in the space-direction interface and the edge modes are described by some CFT. This is consistent with our argument in the physical intuition section. 

\subsection{An Example on Bosonic SPT}
As an example for our result above, we consider the two Ising-paramagnetic systems $\RN{1}$ and $\RN{2}$ as described by Levin and Gu \cite{Levin2012braiding}. The ground state of system $\RN{1}$ is the superposition of all states with different spin configurations. Each spin configuration corresponds to one domain wall (DW) configuration. So we could represent the ground state by the superposition of DW configurations,
\begin{equation}
\ket{\Psi^{\RN{1}}}=\frac{1}{\sqrt{\mathcal N}}\sum_{\{s_1,\dots, s_N\}}\ket{l\text{ DW's}}
\end{equation}
where the normalization factor $\mathcal N=\sum_{\{s_1,\dots, s_N\}}1^l=2^N$ is the total number of spin configurations. This wave function is clearly topologically trivial since it can be rewritten as direct product of the spin triplet state $(\ket{\downarrow_i}+\ket{\uparrow_i})/\sqrt 2$ on each lattice site $i$. And system $\RN{2}$ is a $\mathbb Z_2$ SPT system. The ground-state wave function is the same superposition as system $\RN{1}$ except a factor $(-1)^l$ in front of each spin configuration where $l$ is the number of DW's
\begin{equation}
\ket{\Psi^{\RN{2}}}=\frac{1}{\sqrt{\mathcal N}}\sum_{\{s_1,\dots, s_N\}}(-1)^l\ket{l\text{ DW's}}
\end{equation}
Since states with different spin configurations are mutually orthogonal, the overlap of the normalized wave functions $\ket{\Psi^{\RN{1}}}, \ket{\Psi^{\RN{2}}}$ is
\begin{equation}
\begin{aligned}
\braket{\Psi^{\RN{1}}|\Psi^{\RN{2}}}=&\frac{\sum_{\{s_1,\dots, s_N\}}(-1)^l}{\mathcal N}=\frac{Z_{O(n)}|_{x=1,n=-1}}{\mathcal N/2}\\
Z_{O(n)}=&\sum_{\text{DW config.}} x^L n^l
\end{aligned}
\label{eq: overlap_Levin_Gu}
\end{equation}
where the factor 2 in the denominator is due to the fact that each domain wall configuration corresponds to two different spin configurations. We recognize that the numerator $Z_{O(n)}$ is the partition function of the classical $O(n)$-loop model and $x=1,n=-1$ lies in the critical region \cite{diFrancesco1987modular} with central charge $c=-7$ \cite{diFrancesco1987modular,diFrancesco1987relations, Blote2012the} (See Appendix \ref{appendix: central}). So the finite-size scaling of the numerator follows from Eq. (\ref{eq: CFT_scaling}) with $c=-7$. The denominator $\mathcal N/2$ in Eq. (\ref{eq: overlap_Levin_Gu}) only modifies the non-universal coefficient $\alpha$. Therefore the scaling of $-\ln\braket{\Psi^{\RN{1}}|\Psi^{\RN{2}}}$ is the same as Eq. (\ref{eq: BSPT_scaling}).

\section{Fermionic SPT}
Similar to bosonic SPT, a large class of fermionic SPT phases could be classified using a (special) group supercohomology theory \cite{Gu2014symmetry}. In this paper, we only focus on $(2+1)$-D systems where gapless edge states exist.

\subsection{Ground-state Wave Functions}
The fermionic SPT ground-state wave functions could be constructed in the same way as the bosonic SPT. Suppose the fermionic system with full symmetry $G_f$ is defined on a closed 2-dimensional manifold $M$. The bosonic part of the symmetry is $G_f = G/ \mathbb Z_2^f$ where $\mathbb Z_2^f$ is the fermion-number-parity symmetry. We can extend the hollow 2-dimensional manifold $M$ to solid $(2+1)$-D manifold $\Sigma$ with boundary $\partial \Sigma = M$. Triangulate $M$ and assign a group element $g\in G_b$ to each vertex (see Fig. \ref{fig: triangulation}). Following the same physical intuition as the bosonic case, we can then write down the ground state as the time evolution of arbitrary state from the point $*$ inside $\Sigma$ to the boundary $M$.

In the bosonic SPT systems, a single tetrahedron with vertices 0, 1, 2, 3 is denoted as (0123). We assigned a pure phase (3-cocycle) $\nu_3^{s_{0123}}(g_0, g_1, g_2, g_3)$ where $s_{0123}$ depends the orientation of the tetrahedron $(0123)$. Here, each tetrahedron is associated with a Grassmann tensor. Comparing with the $(d+1)$-D bosonic SPT, the new ingredient is to multiply a Grassmann number for each $d$-simplex. In our interested case of $(2+1)$-D, this means adding a Grassmann number on each triangular face of the tetrahedron. Such a Grassmann tensor in positive oriented tetrahedron is given by
\begin{equation}
\begin{aligned}
\mathcal V_3^+(g_0, g_1, g_2, g_3) =& \nu_3^+(g_0, g_1, g_2, g_3) \theta^{n_2(g_1, g_2, g_3)}_{123}\\
&\theta^{n_2(g_0, g_1, g_3)}_{013}\bar{\theta}^{n_2(g_0, g_2, g_3)}_{023}\bar{\theta}^{n_2(g_0, g_1, g_2)}_{012}
\end{aligned}
\end{equation}
where $\nu_3^+(g_0, g_1, g_2, g_3) =(-1)^{m_1(g_0, g_2)} \nu_3(g_0, g_1, g_2, g_3)$ with $m_1(g_0, g_2) = 0, 1$. The ordering of Grassmann number $\theta$ is naturally inherited from the ordering of the missing indices $0, 2, 4,\dots$ and the ordering of $\bar{\theta}$ is from the missing indices $1, 3, 5, \dots$. The exponent $n_2(g_i, g_j, g_k) = 0, 1$ satisfies
\begin{equation}
\sum_{i=0}^3 n_2(g_0,\dots, \hat{g}_i, \dots, g_3) = \text{even},
\end{equation}
so that the total Grassmann number is even. There is also a relation between $n_2$'s and $m_1$'s,
\begin{equation}
n_2(g_1, g_2, g_3)=m_1(g_2, g_3) + m_1(g_1, g_3) + m_1(g_1, g_2) \text{ mod } 2.
\end{equation}
Such a construction of the sign conventions will be essential in calculating the fermionic path integral in the following.

\begin{figure}[h]
\centering
\includegraphics[scale=0.2]{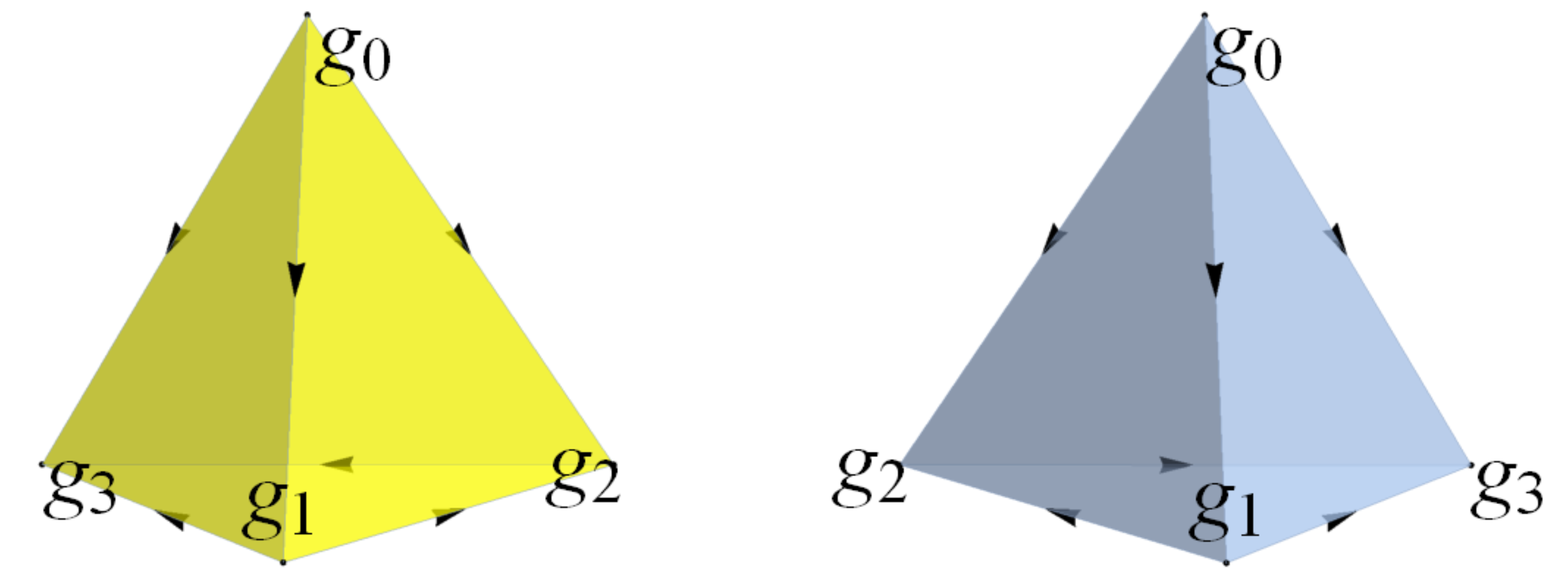}
\caption{Two branched simplexes with opposite orientations. (Left) A simplex with positive orientation; (Right) A simplex with negative orientation.}
\label{fig: Grassmann_tensor}
\end{figure}

For a corresponding negatively oriented tetrahedron [see Fig. \ref{fig: Grassmann_tensor}(b)], the general rule to write down the Grassmann tensor is to reverse the order of $\theta$'s and $\bar{\theta}$'s, and then switch $\theta$ with $\bar{\theta}$ so that $\theta$'s are in front of $\bar{\theta}$'s. As an example, the Grassmann tensor after reversing the orientation of the above tetrahedron $(0123)$ would be
\begin{equation}
\begin{aligned}
\mathcal V_3^-(g_0, g_1, g_2, g_3) =& \nu_3^-(g_0, g_1, g_2, g_3)\theta^{n_2(g_0, g_1, g_2)}_{012}\\
&\theta^{n_2(g_0, g_2, g_3)}_{023}\bar{\theta}^{n_2(g_0, g_1, g_3)}_{013}\bar{\theta}^{n_2(g_1, g_2, g_3)}_{123},
\end{aligned}
\end{equation} 
where $\nu_3^-(g_0, g_1, g_2, g_3) =(-1)^{m_1(g_1, g_3)}/\nu_3(g_0, g_1, g_2, g_3)$.

With each tetrahedron assigned with a Grassmann tensor defined above, we can write down the partition function in terms of fermionic path integral as
\begin{equation}
Z = \frac{1}{|G_b|}\sum_{g_*}\int_{\text{in}(\Sigma)}\prod_{(*ijk)} \mathcal V_3^{s_{ijk}}(g_*, g_i, g_j, g_k)
\end{equation}
where $*$ is the vertex inside the manifold $\Sigma$ and $(*ijk)$ is the 3-simplexes (tetrahedrons) in the triangulation of $\Sigma$. $\text{in}(\Sigma)$ means inside $\Sigma$ but not on the boundary $M$, i.e., $\int_{\text{in}(\Sigma)}$ is a shorthand notation for integral over all Grassmann variables inside $\Sigma$ (up to a sign factor). Explicitly,
\begin{equation}
\begin{aligned}
\int_{\text{in}(\Sigma)} = &\int \prod_{[*ij]}d\theta_{*ij}^{n_2(g_*, g_i, g_j)} d\bar{\theta}_{*ij}^{n_2(g_*, g_i, g_j)} \prod_{\{* i\}} (-1)^{m_1(g_*, g_i)}
\end{aligned}
\label{eq: shorthand}
\end{equation}
where the product $\prod_{\{* i\}}$ is over all the edges $\{*i\}$ connecting the inside vertex $*$ and vertices $i$ on the boundary $M$. Similar to the bosonic case, the fixed-point ground state for the fermionic SPT system is
\begin{equation}
\begin{aligned}
&\Psi(\{g_i\}, \{\theta_{ijk}\}, \{\bar{\theta}_{ijk}\})\\
=& \frac{1}{|G_b|}\sum_{g_*}\int_{\text{in} (\Sigma)} \prod_{(*ijk)} \mathcal V_3^{s_{ijk}}(g_*, g_i, g_j, g_k)\\
=&\int \prod_{[*ij]} d\theta_{*ij}^{n_2(g_*, g_i, g_j)} d\bar{\theta}_{*ij}^{n_2(g_*, g_i, g_j)}\\
&\prod_{\bigtriangledown} \mathcal \nu_3^{-1}(g_*, g_i, g_j, g_k) \prod_{\bigtriangleup} \mathcal \nu_3 (g_*, g_i, g_j, g_k)\\
&\prod_{\bigtriangledown}\theta_{*ij}^{n_2(g_*, g_i, g_j)}\theta_{*jk}^{n_2(g_*, g_j, g_k)} \bar{\theta}_{*ik}^{n_2(g_*, g_i, g_k)} \bar{\theta}_{ijk}^{n_2(g_i, g_j, g_k)}\\
&\prod_{\bigtriangleup}\theta_{ijk}^{n_2(g_i, g_j, g_k)}\theta_{*ik}^{n_2(g_*, g_i, g_k)} \bar{\theta}_{*jk}^{n_2(g_*, g_j, g_k)} \bar{\theta}_{*ij}^{n_2(g_*, g_i, g_j)}
\end{aligned}
\end{equation}
where in the second line we omitted a sign factor \cite{Gu2014symmetry} which does not influence the overlap calculation. $\prod_{[*ij]}$ is the product over all the edges on the triangulation of $M$, $\bigtriangleup$ represents positively oriented triangles and $\bigtriangledown$ represents negatively oriented triangles.

To write down the Hermitian conjugate of the ground-state wave function described above, we need to first understand its triangulation and configuration. Similar to the bosonic case, we could take the mirror image of the triangulation for the original system. Denote the mirror image of $\Sigma$ as $\widetilde{\Sigma}$ and the point inside $\widetilde{\Sigma}$ as $\#$. Then reverse the arrow between the point $\#$ and all of the vertices on $\widetilde{M}$ so that all arrows point to $\#$. This new configuration has the same interpretation as time evolution of arbitrary initial state from $+\infty$ to 0. The resulting wave function corresponds to the Hermitian conjugate of $\Psi$. Mathematically,
\begin{equation}
\begin{aligned}
&\Psi^\dagger(\{g_i\}, \{\theta_{ijk}\}, \{\bar{\theta}_{ijk}\})\\
=&\int \prod_{[ij\#]} d\theta_{ij\#}^{n_2(g_i, g_j, g_\#)} d\bar{\theta}_{ij\#}^{n_2(g_i, g_j, g_\#)}\\
&\prod_{\bigtriangleup} \mathcal \nu_3^{-1}(g_i, g_j, g_k, g_\#) \prod_{\bigtriangledown} \mathcal \nu_3 (g_i, g_j, g_k, g_\#)\\
&\prod_{\bigtriangleup}\theta_{ij\#}^{n_2(g_i, g_j, g_\#)}\theta_{jk\#}^{n_2(g_j, g_k, g_\#)} \bar{\theta}_{ik\#}^{n_2(g_i, g_k, g_\#)} \bar{\theta}_{ijk}^{n_2(g_i, g_j, g_k)}\\
&\prod_{\bigtriangledown}\theta_{ijk}^{n_2(g_i, g_j, g_k)}\theta_{ik\#}^{n_2(g_i, g_k, g_\#)} \bar{\theta}_{jk\#}^{n_2(g_j, g_k, g_\#)} \bar{\theta}_{ij\#}^{n_2(g_i, g_j, g_\#)}.
\end{aligned}
\end{equation}
Here $\bigtriangleup, \bigtriangledown$ represent negatively and positively oriented triangles in the mirror image $\widetilde{M}$ respectively, exactly opposite to the formula in the $\Psi$ configuration.

\subsection{Overlaps between Fermionic SPT Ground States}
Calculating the overlap of $\Psi$ and its Hermitian conjugate is equivalent to gluing the $(2+1)$-D complexes $\Sigma$ and $\widetilde{\Sigma}$ over their common boundary $M$ (or the mirror image $\widetilde{M}$)\cite{Gu2014symmetry}. In general, the overlap of two ground states $\ket{\Psi^{\RN{1}}}, \ket{\Psi^{\RN{2}}}$ is defined as
\begin{equation}
\begin{aligned}
&\braket{\Psi^{\RN{1}}|\Psi^{\RN{2}}} = \tr\left(\Psi^{\RN{2}} (\Psi^{\RN{1}})^\dagger\right)= \frac{1}{|G_b|^{N+2}}\sum_{\{\text{all }g\text{ on } \Sigma \cup \widetilde{\Sigma}\}}\\
&\int_M \left(\int_{\text{in}(\widetilde{\Sigma})} \prod_{[ijk]}(\mathcal V^{\RN{2}}_3)^{s_{ijk}} \int_{\text{in}(\Sigma)} \prod_{[ijk]}(\mathcal V^{\RN{1}}_3)^{s_{ijk}}\right)
\end{aligned}
\end{equation}
where $N$ is the number of lattice sites on the manifold $M$, $|G_b|$ is the order of group $G_b$, and $M = \Sigma \cap \widetilde{\Sigma}$ is the common boundary of $\Sigma$ and $\widetilde{\Sigma}$. In this equation, we follow the shorthand notation defined in Eq. (\ref{eq: shorthand}).

Similar to the bosonic case, the overlap of $\Psi$ with its Hermitian conjugate is a fermionic path integral over a closed manifold $\Sigma \cup \widetilde{\Sigma}$ and expected to be 1. Indeed, it was explicitly shown by Gu and Wen \cite{Gu2014symmetry} that on any $(2+1)$-D closed manifold the partition function reduces to
\begin{equation}
\int \mathcal V_3^+(g_0, g_1, g_2, g_3) \mathcal V_3^-(g_0, g_1, g_2, g_3) = 1
\end{equation}
where $\int$ represents the integration over all Grassmann variables up to a sign factor. This result naturally leads to $\braket{\Psi|\Psi} = 1$, consistent with the normality of quantum states.

From the construction of Grassmann tensors we clearly see that the sign factors $n_2$'s and $m_1$'s only depend on the symmetry $G_f$ and triangulation. The only factor encoding the topological theory is the cocycles $\nu_3$. So if we have two different SPT phases defined on the same manifold $M$, the Grassmann integral part of the overlap $\braket{\Psi^{\RN{1}}|\Psi^{\RN{2}}}$ would be the same as calculating $\braket{\Psi^{\RN{1}}|\Psi^{\RN{1}}}$ or $\braket{\Psi^{\RN{2}}|\Psi^{\RN{2}}}$, and thus contributes 1 to the path integral. This was also pointed out by Gu and Wen \cite{Gu2014symmetry}. They found that on the closed manifolds the integrals over Grassmann tensor give rise to complex numbers $e^{-S}$. So according to their results all the integrals over Grassmann numbers should also contribute 1, consistent with our analysis. After canceling out all the Grassmann numbers, the fermionic path integral $\braket{\Psi^{\RN{1}}|\Psi^{\RN{2}}}$ leads to the same bosonic path integral expression as in the Bosonic case [Eq. (\ref{eq: bosonic_SPT})]. Therefore the same argument leads to a WZW theory defined on the boundary $M$. So the amplitude of the overlap is again of the form
\begin{equation}
\ln \braket{\Psi^{\RN{1}}|\Psi^{\RN{2}}}= -\alpha N+\frac{\chi c}{12}\ln N+O(1).
\end{equation}

\section{Intrinsic Topological Orders}
In this section, we use the famous FQH states to show that the term $\chi c/12 \ln(N)$ still exists in the scaling of ground-state overlaps. Besides that, we also find a leading term decaying faster than exponential. This is quite surprising since traditional view would expect exponential decay as leading term. In the following we will show the rigorous calculation.

\subsection{FQH Wave Functions on Disks}
FQH states have been well-known to exhibit intrinsic topological order \cite{Wen1989vacuum}. Among these states, $1/m$ filling FQH states can be described by Laughlin wave functions \cite{Laughlin1983anomalous}, which are given on the disk as
\begin{equation}
\Psi_D^{\RN{2}}=\frac{1}{\sqrt{Z_d}}\prod_{i<j} (z_i-z_j)^m e^{-\sum_k \frac{|z_k|^2}{4}}
\label{eq: Laughlin_wave_function}
\end{equation}
where $Z_d$ is the normalization factor on the disk
\begin{equation}
Z_d=\int_0^R d^2z_1\cdots \int_0^R d^2 z_N e^{-\frac{1}{2} \sum_i |z_i|^2} \prod_{i<j} |z_i-z_j|^{2m}.
\end{equation}

Due to rotational symmetry, the many-body angular momentum $J=m N(N-1)/2$ is a good quantum number and wave functions with different filling or particle number live in distinct Hilbert space. The meaningful overlaps should be calculated between Laughlin wave functions and topologically trivial wave functions with the same particle number, filling factor, and hence the same angular momentum (\ref{eq: Laughlin_wave_function}). To find such a trivial wave function, we note that Laughlin wave functions can be expanded to a series of Slater determinants $\mathcal D_\lambda$ \cite{Dunne1993slater}.
\begin{equation}
\begin{aligned}
\Psi_D^{\RN{2}}&=\frac{1}{Z_d}\sum_\lambda a_\lambda \mathcal D_\lambda,\\
\mathcal D_\lambda &=\begin{vmatrix}
z_1^{\lambda_1} & z_1^{\lambda_2} & \cdots & z_1^{\lambda_N}\\
z_2^{\lambda_1} & z_2^{\lambda_2} & \cdots & z_2^{\lambda_N}\\
\vdots & \vdots & \vdots & \vdots\\
z_N^{\lambda_1} & z_N^{\lambda_2} & \cdots & z_N^{\lambda_N}
\end{vmatrix}
\exp \left(-\sum_{i=1}^N \frac{|z_i|^2}{4} \right)
\end{aligned}
\end{equation}
where $\lambda=(\lambda_1,\lambda_2, \dots, \lambda_N)$ is the partition of $J=m N(N-1)/2$. Each term $\mathcal D_\lambda$ in the expansion is a topologically trivial direct-product state with the same filling fraction and particle number as the original Laughlin wave function. And all of $\mathcal D_\lambda$'s constructed in this way are mutually orthogonal (See Appendix \ref{appendix: slater}). We choose the Slater determinant $\mathcal D_{\lambda^{(1)}}$ with $\lambda^{(1)}=(m(N-1), m(N-2),\dots, 0)$ as the topologically trivial wave function since its coefficient $a_{\lambda^{(1)}}=1$. So the normalized trivial wave function is
\begin{equation}
\begin{aligned}
\Psi_D^{\RN{1}}=&\frac{1}{Z_d^{(1)}}\mathcal D_{\lambda^{(1)}},\\
Z_d^{(1)}=&\int_0^R d^2z_1\cdots \int_0^R d^2 z_N e^{-\frac{1}{2} \sum_i |z_i|^2} \prod_{i<j} |z_i^m-z_j^m|^2.
\end{aligned}
\end{equation}

Then the overlap of $\Psi_D^{\RN{1}}$ and $\Psi_D^{\RN{2}}$ is
\begin{equation}
\braket{\Psi_D^{\RN{1}}|\Psi_D^{\RN{2}}}=\frac{\sum_\lambda a_\lambda \braket{\mathcal D_{\lambda^{(1)}}|\mathcal D_\lambda}}{\sqrt{Z_d^{(1)}Z_d}}=\sqrt{\frac{Z_d^{(1)}}{Z_d}}=\sqrt{\frac{A(N) Z_d^{(1)}}{Z_D}}.
\label{eq: disk_overlap}
\end{equation}
where the denominator $Z_D=A(N) Z_d$ is the partition function of the one component plasma (OCP). And the prefactor $A(N)$ (See Appendix \ref{appendix: one_disk}) comes from the ideal gas partition function, the background-background interaction and the constant part of particle-background interaction
\begin{equation}
A(N)=\frac{1}{N!}\left(\frac{\pi m M}{h^2}\right)^N L^{m N}e^{\frac{3m N^2}{4}} (2m N)^{-\frac{m N^2}{2}}.
\end{equation}
where $L$ is an arbitrary length scale, $M$ is the mass of each charged particle and $h$ is the Planck constant. Here we have also converted the disk radius $R$ into the particle number $N$ since the particle density is fixed at $n= N/(\pi R^2)\equiv 1/(2\pi m)$ in the plasma analogy \cite{Laughlin1983anomalous}.

The OCP can be considered as a critical system with central charge $c=-1$ \cite{Jancovici1994coulomb, Jancovici1996coulomb, Tellez1999exact, Jancovici2000universal}. So according to Eq. (\ref{eq: CFT_scaling}) the finite-size scaling of the partition function $Z_D$ on a disk is
\begin{equation}
\ln Z_D=-\alpha_D N-\beta_D \sqrt N-\frac{1}{12} \ln N+O(1)
\end{equation}
where we used $\chi=1$ for the disk. Following the method of Caillol \cite{Caillol1981exact}, $Z_d^{(1)}$ in the numerator could be solved exactly by expanding the polynomial into a summation of monomials and then transforming the integral to polar coordinates. Integrating out the angular part of each coordinate we find that all terms with nonzero phases vanish.
\begin{equation}
\begin{aligned}
Z_d^{(1)}=&(2\pi)^N N!\prod_{i=1}^N\int_0^R r_idr_i \prod_{j=1}^N r_j^{m(j-1)} e^{-\frac{1}{2} \sum_k r_i^2}\\
=&(2\pi)^N 2^{\frac{m(N-1)N}{2}} \prod_{j=1}^N [m (j-1)]! N! Z_{\text{bdry}}^{(1)}\\
Z_{\text{bdry}}^{(1)}&=\prod_{i=1}^N \left ( \frac{1}{ [m (i-1)]!} \int_0^{m N} dx_i e^{-x_i} x_i^{m(i-1)}\right )
\end{aligned}
\end{equation}
where we made the coordinate transformation $x_i=r_i^2/2$. Those terms in $Z_{\text{bdry}}^{(1)}$ are close to 1 if $i$ is small. So the only terms that contribute dramatically to $Z_{\text{bdry}}^{(1)}$ are those $i$'s of order $\sqrt N$ or greater. Then the following asymptotic formula holds \cite{dlmf_nist_gov}
\begin{equation}
\begin{aligned}
&\frac{1}{ [m(i-1)]!} \int_0^{m N} dx_i e^{-x_i} x_i^{m(i-1)}\\
=&\frac{1}{2} \left [1+\erf \left(\frac{m(N-i+1)}{\sqrt{2m N}}\right) \right ]+O \left(\frac{1}{\sqrt N}\right)
\end{aligned}
\end{equation}
where $\erf(x)=\frac{2}{\sqrt \pi} \int_0^x e^{-t^2} dt$ is the error function. Replacing the summation over $i$ by an integral over $y=m(N-i+1)/\sqrt{2m N}$, we find $\ln Z_{\text{bdry}}^{(1)}$ scales as $\sqrt N$
\begin{equation}
\ln Z_{\text{bdry}}^{(1)}=\sqrt{\frac{2N}{m}} \int_0^\infty dy \ln\left(\frac{1+\erf (y)}{2}\right)+O(1).
\end{equation}
The terms with factorials could be evaluated through converting the summations into integrals using Euler-Maclaurin formula (See Appendix \ref{appendix: scaling_disk}). Putting every term together, the finite-size scaling for the overlap is
\begin{equation}
\ln \braket{\Psi_D^{\RN{1}}|\Psi_D^{\RN{2}}}=-a N \ln N-\alpha N -\beta \sqrt N+\gamma \ln N+O(1)
\end{equation}
where $a=\frac{m-1}{4}, \alpha=\frac{1}{4}[(m-1) (\ln m-1)+m \ln 2-3 \ln(2\pi)-2 \ln(\frac{m M\pi}{h^2}) -2\alpha_D], \beta=-\frac{1}{2}[\sqrt{\frac{2}{m}} \int_0^\infty dy \ln(\frac{1+\erf (y)}{2})+\beta_D], \gamma=\frac{(m-1)^2}{24m}$.

\subsection{FQH Wave Functions on 2-spheres}
To show the effect of the topology, we also calculated the overlap on a sphere. Similar to the disk case, the $N$-particle Laughlin wave functions on the sphere can also be decomposed into Slater determinants $\mathcal S_\lambda$. And we choose the trivial wave function as the one with $\lambda^{(1)}=(m(N-1),m(N-2),\dots,0)$.
\begin{equation}
\begin{aligned}
\Psi_{S^2}^{\RN{2}}=&\frac{1}{Z_s}\prod_{i<j} (u_i v_j-u_j v_i)^m\\
=&\frac{1}{Z_s}\prod_{k=1}^N u_k^{m(N-1)}\prod_{i<j}\left(\frac{v_j}{u_j}-\frac{v_i}{u_i}\right)^m=\frac{1}{Z_s} \sum_\lambda b_\lambda \mathcal S_\lambda,\\
\Psi_{S^2}^{\RN{1}}=&\frac{1}{Z_s^{(1)}} \mathcal S_{\lambda^{(1)}}=\frac{1}{Z_s^{(1)}}\prod_{k=1}^N u_k^{m(N-1)}\prod_{i<j}\left[\left(\frac{v_j}{u_j}\right)^m-\left(\frac{v_i}{u_i}\right)^m\right],
\end{aligned}
\end{equation}
where $u_i=\cos\left(\frac{\theta_i}{2}\right)e^{i\phi_i/2}, v_i=\sin\left(\frac{\theta_i}{2}\right)e^{-i\phi_i/2}$ are the spinor coordinates. And the normalization factors are
\begin{equation}
\begin{aligned}
Z_s=&\int d\Omega_1 \cdots d\Omega_N \prod_k |u_k|^{2m (N-1)} \prod_{i<j} \left |\frac{v_j}{u_j}-\frac{v_i}{u_i} \right|^{2m},\\
Z_s^{(1)}=&\int d\Omega_1 \cdots d\Omega_N \prod_k |u_k|^{2m (N-1)} \\
&\prod_{i<j} \left |\left(\frac{v_j}{u_j}\right)^m-\left(\frac{v_i}{u_i}\right)^m \right|^2.
\end{aligned}
\end{equation}

The same argument as in the disk case shows
\begin{equation}
\begin{aligned}
\braket{\Psi_{S^2}^{\RN{1}}|\Psi_{S^2}^{\RN{2}}}=&\sqrt{\frac{Z_s^{(1)}}{Z_s}}=\sqrt{\frac{B(N)Z_s^{(1)}}{Z_{S^2}}},\\
Z_{S^2}=&B(N)Z_s,\\
B(N)=\frac{1}{N!}&\left(\frac{m M\pi}{h^2}\right)^N \left(\frac{L}{2}\right)^{mN} \left(\frac{mN}{2}\right)^{\frac{(2-m)N}{2}} e^{mN^2/2}
\label{eq: sphere_overlap}
\end{aligned}
\end{equation}
where $B(N)$ (See Appendix \ref{appendix: one_sphere}) comes from the partition function of $N$-particle ideal gas, background-background interaction, the constant part of particle-background interaction and the radius dependence of the integral in the partition function of the OCP.

As in the disk case, the denominator $Z_{S^2}$ is the partition function of the OCP on a sphere and scales as
\begin{equation}
\ln Z_{S^2}=-\alpha_{S^2} N-\frac{1}{6}\ln N+O(1)
\end{equation}
where we used $\chi=2$ for spheres. And the $\sqrt N$ term vanishes because spheres do not have boundaries.

To evaluate $Z_s^{(1)}$ in the numerator, we follow Alastuey and Jancovici \cite{Alastuey1981on} to change variables $r_i=\tan\left(\frac{\theta_i}{2}\right)$ or $\sin \theta_i d\theta_i=\frac{4r_i dr_i}{(1+r_i^2)^2}$. Then the integral is over the plane (with no boundaries) defined by the polar coordinates $(r_i, \phi_i)$. Again, the only terms contributing to the integral are those terms with vanishing polar angles.
\begin{equation}
\begin{aligned}
Z_s^{(1)}=&N! (2\pi)^N \prod_i \int \frac{4r_i^{2m(i-1)+1} dr_i}{(1+r_i^2)^{m(N-1)+2}}\\
=&N! (4\pi)^N \prod_i \frac{[m(i-1)]![m(N-i)]!}{[m(N-1)+1]!}
\end{aligned}
\label{eq: first_partition_in_expansion}
\end{equation}
Taking the logarithm of both sides and then using Euler-Maclaurin formula to convert the summation to integral (See \ref{appendix: scaling_sphere}), we find the scaling of the overlap to be
\begin{equation}
\ln \braket{\Psi_{S^2}^{\RN{1}}|\Psi_{S^2}^{\RN{2}}}=-a N\ln N-\alpha N+\gamma \ln N+O(1)
\label{eq: sphere_scaling}
\end{equation}
where $a=\frac{m-1}{4},\alpha=\frac{1}{4}[(m-1) [\ln (2m)-2]-2m \ln L-2 \ln(\frac{m M\pi}{h^2})-2\alpha_{S^2}],\gamma=\frac{(m-1)^2}{12m}$ and the boundary term $\sqrt N$ vanishes explicitly.

Comparing the results for disks and spheres, we find that the leading order coefficients for both cases are the same $a=(m-1)/4$. Such a term indicates a faster decay than typical Anderson orthogonality catastrophe. The coefficient of $\ln N$ is proportional to the Euler characteristic of the manifold where the system is defined. This term stems from the critical behavior of the classical OCP and thus is consistent with our physical intuition pictured in the introduction section.

\section{Conclusions and Open Questions}
In this paper, we calculated the finite-size scaling of the overlaps of topologically different states through mapping the overlap to partition functions of critical systems. For generic $(2+1)$-D topologically different states, including both SPT states and intrinsic topological states, the fixed-point ground-state overlaps exhibit a universal sub-leading term $\frac{\chi c}{12}\ln N$ depending on the topology of the manifold on which the system lives. Such a universal topological response term relies on the gapless edge state on the interface of different topological systems, as described in Fig. \ref{fig: physical_picture}. In $(2+1)$-D, SPT systems are known to have gapless edge modes described by some CFT. So we conclude that the topological response term always exists for $(2+1)$-D SPT systems at fixed-points. Besides the CFT describing the \textit{space-direction} edge modes, the central charge $c$ indicates that there is another CFT associated with the two systems in the \textit{time-direction} interface (see Fig. \ref{fig: physical_picture}). It has been clear to us that this CFT from the overlap calculation is different from the CFT on the space-direction interface between the two systems. But it seems to appear when gapless edge modes exist on the space-direction interface of the two systems. We \textit{conjecture} that the two CFT's are related in some way. Maybe the Wick rotation of a CFT in the time direction interface becomes the CFT on the interface of space direction (see Fig. \ref{fig: physical_picture}).

In the case of intrinsic topological order, we only calculate the overlaps between the famous FQH states and product states. We find that the same topological response term still exists and it is also derived from a critical system. Thus we may \textit{conjecture} that the topological response term persists for any $(2+1)$-D topological systems with gapless edge modes. In this overlap calculation, we also notice a leading-order scaling $-\frac{m-1}{4} N \ln N$ which decays faster than the expected exponential in typical Anderson orthogonality catastrophe. Its coefficient only depends on which Laughlin state participates in the overlap calculation. Such a behavior may be used as a signature for topological phase transitions between topologically ordered FQH states and trivial states.

In higher dimensions, the boundaries of SPT systems may also be gapped by symmetry breaking or due to intrinsic surface topological order \cite{Vishwanath2013physics, Burnell2014exactly, Fidkowski2013nonabelian, Chen2014symmetry, Metlitski2015symmetry}. In such systems, the overlap can not have $\ln N$ correction. But there might be other universal terms such as a constant term similar to the subleading term in entanglement entropy. We will leave this as future work.

\begin{acknowledgments}
J. Gu would like to thank Professor Kai Sun for his advice and support throughout this project, Ying-Hai Wu for discussions on the expansion of Laughlin wave functions, Huan He and Di Zhou for their help in the beginning of this project. This work was supported by the National Science Foundation (NSF grant EFRI-1741618).
\end{acknowledgments}

\appendix

\section{Central charge of the critical $O(n)$-loop model \label{appendix: central}}

The partition function of the $O(n)$-loop model on the honeycomb lattice is
\begin{equation}
Z_{O(n)}=\sum_{\text{DW config.}} x^L n^l
\end{equation}
for general values of $x$ and $n\in[-2,2]$. The critical line \cite{Nienhuis1982exact} of this model is $x_c=\left[2+(2-n)^{1/2}\right]^{-1/2}$. For $x>x_c$ which contains the point $x=1, n=-1$, this loop model is also critical \cite{diFrancesco1987modular}. And the central charges of both cases are given by \cite{diFrancesco1987modular,diFrancesco1987relations}
\begin{equation}
c=1-\frac{6(g-1)^2}{g}
\label{eq: central_charge}
\end{equation}
where $g$ is defined by $n=-2\cos(\pi g)$. The branches $g\in [0,1]$ and $g\in [1,2]$ correspond to $x>x_c$ and $x=x_c$ systems respectively. The central charge for $x=1,n=-1$ is found through Eq. (\ref{eq: central_charge}) by setting $g=1/3$, which leads to $c=-7$ (numerically verified by Bl\"ote et al \cite{Blote2012the}).

Actually, the denominator $\mathcal N/2$ of the overlap (\ref{eq: overlap_Levin_Gu}) of the main text can also be considered trivially as a critical $O(n)$-loop model at $x=1, n=1$. Then the same formula gives $c=0$, implying no logarithmic term correction in the ``free energy''. This is consistent with $\mathcal N/2=2^{N-1}$ we got from direct counting.

\section{Slater determinants from Laughlin wave-function expansion are mutually orthogonal \label{appendix: slater}}

A typical Slater determinant in the expansion of Laughlin wave functions is as follows,
\begin{equation}
\begin{aligned}
\mathcal D_{\lambda}=&\begin{vmatrix}
z_1^{\lambda_1} & z_1^{\lambda_2} & \cdots & z_1^{\lambda_N}\\
z_2^{\lambda_1} & z_2^{\lambda_2} & \cdots & z_2^{\lambda_N}\\
\vdots & \vdots & \vdots & \vdots\\
z_N^{\lambda_1} & z_N^{\lambda_2} & \cdots & z_N^{\lambda^N}
\end{vmatrix}
\exp \left(-\sum_{i=1}^N \frac{|z_i|^2}{4} \right)\\
=&\sum_{\sigma\in S^N} \sgn(\sigma)z_1^{\sigma(\lambda_1)}z_2^{\sigma(\lambda_2)}\cdots z_N^{\sigma(\lambda_N)}
e^{-\frac{1}{4} \sum_i r_i^2}
\end{aligned}
\end{equation}
where $\lambda=(\lambda_1,\lambda_2,\dots, \lambda_N)$ is a partition of $J=m N(N-1)/2$. Another partition $\mu\neq \lambda$ corresponds to a different Slater determinant in the expansion. Their inner product is given by
\begin{widetext}
\begin{equation}
\begin{aligned}
&\int \prod_{i=1}^N dz_i \mathcal D_\lambda^{(1)} D_\mu=\int \prod_{i=1}^N dz_i \left[\sum_{\sigma_\lambda\in S^N} \sgn(\sigma_\lambda)(z_1^*)^{\sigma_\lambda(\lambda_1)}(z_2^*)^{\sigma_\lambda(\lambda_2)}\cdots (z_N^*)^{\sigma_\lambda(\lambda_N)}
e^{-\frac{1}{4} \sum_i r_i^2}\right]\\
&\left[\sum_{\sigma_\mu\in S^N} \sgn(\sigma_\mu)z_1^{\sigma_\mu(\mu_1)}z_2^{\sigma_\mu(\mu_2)}\cdots z_N^{\sigma_\mu(\mu_N)}
e^{-\frac{1}{4} \sum_i r_i^2}\right]\\
=&\int \prod_{i=1}^N (r_i dr_i d\phi_i) \sgn(\sigma_\lambda \sigma_\mu)r_1^{\sigma_\lambda(\lambda_1)+\sigma_\mu(\mu_1)}r_2^{\sigma_\lambda(\lambda_2)+\sigma_\mu(\mu_2)}\cdots r_N^{\sigma_\lambda(\lambda_N)+\sigma_\mu(\mu_N)}\\
&e^{i(\sigma_\mu(\mu_1)-\sigma_\lambda(\lambda_1))\phi_1}e^{i(\sigma_\mu(\mu_2)-\sigma_\lambda(\lambda_2))\phi_2}\cdots e^{i(\sigma_\mu(\mu_N)-\sigma_\lambda(\lambda_N))\phi_N}e^{-\frac{1}{2} \sum_i r_i^2}
\end{aligned}
\end{equation}
\end{widetext}
Clearly, the integral over polar angles vanishes unless $\sigma_\mu(\mu_i)-\sigma_\lambda(\lambda_i)=0$ for all $i\in \{1,2,\dots,N\}$. But if this condition is true, then $\lambda=\mu$ which contradicts with our assumption that $\lambda$ and $\mu$ are different partitions of $J=mN(N-1)$. Therefore the integral must vanish and all the Slater determinants in the expansion of Laughlin wave functions are mutually orthogonal.

\section{One component plasma on a disk \label{appendix: one_disk}}
This section is based on the paper by Sari et al \cite{Sari1976on}.

One component plasma (OCP) on a disk consists of $N$ identical particles with charge $e$ and a neutralizing background with uniform charge distribution. The Hamiltonian is $H=T+V$ where
\begin{equation}
T=\sum_{i=1}^N \frac{p_i^2}{2M}
\end{equation}
is the kinetic term and the potential term $V$ consists of background-background interaction, particle-background interaction and particle-particle interaction.
\begin{equation}
\begin{aligned}
&V=V_{bb}+V_{pb}+V_{pp}\\
=&-\frac{e^2 n^2}{2} \int d^2 z d^2 w \ln\left|\frac{z-w}{L}\right|+e^2 n \sum_i\int d^2 w \ln\left|\frac{w-z_i}{L}\right|\\
&-\frac{e^2}{2} \sum_{i\neq j} \ln\left|\frac{z_i-z_j}{L}\right|
\end{aligned}
\end{equation}
where $L$ is an arbitrary length scale.

After the integration and using the relation $n=N/(\pi R^2)$ we find
\begin{equation}
\begin{aligned}
V_{bb}=&-\frac{e^2 N^2}{2}\left [\ln \left(\frac{R}{L}\right)-\frac{1}{4} \right ],\\
V_{pb}=&\frac{e^2 N}{2}\sum_i\left[2 \ln\left( \frac{R}{L}\right)-1+\left(\frac{r_i}{R}\right)^2\right]\\
=&\frac{e^2 N^2}{2}\left[2 \ln\left( \frac{R}{L}\right)-1\right]-\frac{e^2 N}{2}\sum_i\left(\frac{r_i}{R}\right)^2.
\end{aligned}
\end{equation}

So
\begin{equation}
\begin{aligned}
V=&-\frac{e^2 N^2}{2} \left[ \frac{3}{4}-\ln\left(\frac{R}{L}\right)\right]+\frac{e^2 N}{2}\sum_i\left(\frac{r_i}{R}\right)^2\\
&-\frac{e^2}{2} \sum_{i\neq j} \ln \left|\frac{z_i-z_j}{L}\right|.
\end{aligned}
\end{equation}

Then the canonical partition function could be calculated by
\begin{equation}
Z=\tr(e^{-\beta H}).
\end{equation}
The kinetic term $T=\sum_i p_i^2/(2M)$ contributes as the partition function of ideal gas. Integrate out the momenta first and we obtain
\begin{equation}
\begin{aligned}
&\frac{1}{N!} \prod_i \int \frac{d^2 z_i d^2p_i}{h^2} \exp\left(-\beta \frac{p_i^2}{2M}\right)\\
=&\frac{1}{N!}\left(\frac{2M\pi}{\beta h^2}\right)^N \prod_i \int d^2 z_i
\end{aligned}
\end{equation}
where we leave the integral over particle positions untouched because the potential part of the plasma depends on positions.

To relate OCP partition function with Laughlin wave function, we assume the particle density $n=\frac{1}{2 \pi m}$, electric charge $e=m$ and $\beta=1/(k_B T)=2/m$. Then the partition function $Z_D$ of the OCP on a disk is
\begin{equation}
\begin{aligned}
Z_D\equiv & A(N) Z_d\\
A(N)=&\frac{L^{m N}}{N!}e^{\frac{3m N^2}{4}} (2m N)^{-m N^2/2} \left(\frac{m \pi M}{h^2}\right)^N\\
Z_d=&\int d^2 z_1\cdots \int d^2 z_N e^{-\frac{1}{2}\sum_i r_i^2}\prod_{i< j} |z_i-z_j|^{2m}
\end{aligned}
\end{equation}
where $Z_d$ is exactly the normalization factor of Laughlin wave function and $A(N)$ is the prefactor independent of the integral.

\section{Scaling of $A(N) Z_d^{(1)}$ in the disk case \label{appendix: scaling_disk}}

To obtain the scaling of $A(N) Z_d^{(1)}$ in the numerator of the overlap function (\ref{eq: disk_overlap}) in the main text, the only difficulty comes from the evaluation of $\prod_{i=1}^N [m(i-1)]!$. We first take logarithm of this term to convert it to a summation $\sum_{i=1}^N \ln [m(i-1)]!$ and then use Euler-Maclaurin formula to change the summation to an integral with a controlled error term. The Euler-Maclaurin formula reads
\begin{equation}
\begin{aligned}
\sum_{i=m+1}^n f(i)=&\int_m^n f(x) dx+\sum_{k=1}^p \frac{B_k}{k!} (f^{(k-1)}(n)-f^{(k-1)}(m))\\
&+\text{higher order error terms}
\end{aligned}
\label{eq: appendix_Euler_Maclaurin}
\end{equation}
where $B_k$ is the $k$-th Bernoulli number and the error term depends on which term $p$ we stop the expansion. For our purpose, knowing $B_1=\frac{1}{2}, B_2=\frac{1}{6}$ is enough.

But before we can finish the calculation, $\ln [m(i-1)]!$ must be evaluated first using Stirling's formula (or Euler-Maclaurin formula)
\begin{equation}
\begin{aligned}
&f(i)=\ln [m(i-1)]!\\
=&m(i-1)\ln [m(i-1)]-m(i-1)+\frac{1}{2} \ln [m(i-1)]\\
&+\frac{1}{2}\ln(2\pi)+\frac{1}{12 m(i-1)}+\cdots
\end{aligned}
\label{eq: appendix_Stirling}
\end{equation}

Taking the derivative of this equation we find
\begin{equation}
\frac{df(i)}{di}=m \ln[m(i-1)]+\frac{1}{2(i-1)}+\cdots.
\end{equation}

Plug these into the Euler-Maclaurin formula, we find
\begin{widetext}
\begin{equation}
\begin{aligned}
&\sum_{i=1}^N \ln [m(i-1)]!-\ln[m!]\\
=&\int_{i=2}^N \left\{m(i-1)\ln [m(i-1)]-m(i-1)+\frac{1}{2} \ln [m(i-1)]+\frac{1}{2}\ln(2\pi)+\frac{1}{12 m(i-1)}\right\}\\
&+\frac{1}{2}\left\{m(N-1)\ln[m(N-1)]-m(N-1)+\frac{1}{2}\ln[m(N-1)]-\ln[m!]\right\}\\
&+\frac{1}{12}\{m \ln[m(N-1)]-m\ln m\}+O(1)\\
=&-\frac{m}{2} N^2\ln N+\frac{2m\ln m-3m}{4}N^2-\frac{m-1}{2}N\ln N-\frac{1}{2}[(m-1)(\ln m-1)-\ln(2\pi)] N\\
&+\frac{m^2-3m+1}{12m}\ln N+O(1).
\end{aligned}
\end{equation}
\end{widetext}
Then it is an easy calculation to find the scaling of the overlap as in the main text.

\section{One component plasma on a sphere \label{appendix: one_sphere}}

Similar to the OCP on a disk, the Hamiltonian of the OCP on a sphere consists of a kinetic term $K=\sum_{i=1}^N p_i^2/(2M)$ and potential term $U$ coming from background-background interaction, particle-background interaction and particle-particle interaction. In the following, the distance between two points on the sphere are calculated by embedding the sphere into a 3D Euclidean space.
\begin{equation}
\begin{aligned}
U=&U_{bb}+U_{pb}+U_{pp}\\
U_{bb}=&-\frac{e^2 n^2}{2}\int R^2 d\Omega_x R^2 d\Omega_y \ln \left|\frac{\boldsymbol x-\boldsymbol y}{L}\right|\\
=&-\frac{e^2 N^2}{4}\left[ 2\ln \left(\frac{2R}{L}\right) -1\right]\\
U_{pb}=&e^2 n \sum_i \int R^2 d\Omega_x \ln\left|\frac{\boldsymbol x_i-\boldsymbol x}{L}\right|\\
=&\frac{e^2 N^2}{2} \left[2\ln\left(\frac{2R}{L}\right)-1\right]\\
U_{pp}=&-e^2 \sum_{i<j}\ln \left |\frac{\boldsymbol x_i-\boldsymbol x_j}{L}\right |\\
=&-e^2 \sum_{i<j} \ln\left|\frac{2R}{L}(u_i v_j-u_j v_i)\right|
\end{aligned}
\end{equation}
where $u=\cos \frac{\theta}{2} e^{i \phi/2}, v=\sin\frac{\theta}{2}e^{-i \phi/2}$.

In the Laughlin plasma analogy, $\beta=2/m$ and $e=m, n=1/(2\pi m)$. So the partition function (including the ideal gas part) is
\begin{equation}
\begin{aligned}
&Z_{S^2}= \tr[e^{-\beta (K+U)}]=\frac{ e^{mN^2/2} }{N!}\left(\frac{m M\pi}{h^2}\right)^N \left(\frac{L}{2R}\right)^{mN^2} \\
&\int R^2 d\Omega_1\cdots R^2 d\Omega_N  \prod_{i<j} \left[ \frac{2R}{L} |u_i v_j-u_j v_i | \right]^{2m}\\
=& \frac{ e^{mN^2/2} }{N!}\left(\frac{m M\pi}{h^2}\right)^N \left(\frac{L}{2}\right)^{mN} \left(\frac{mN}{2}\right)^{\frac{(2-m)N}{2}}  \\
&\int d\Omega_1\cdots d\Omega_N \prod_{k=1}^N |u_k|^{2m (N-1)}\prod_{i<j} \left |\frac{v_j}{u_j}-\frac{v_i}{u_i} \right|^{2m}
\end{aligned}
\end{equation}
Then we recognize that the integral part is the normalization factor $Z_s$ of Laughlin wave functions and the prefactor is $B(N)$.

\section{Scaling of $B(N) Z_s^{(1)}$ in the sphere case \label{appendix: scaling_sphere}}

In the main text, the scaling of factorial part of $Z_s^{(1)}$ in Eq. (\ref{eq: first_partition_in_expansion}) is the only difficulty for finding the scaling of $B(N)Z_s^{(1)}$ in the overlap (\ref{eq: sphere_overlap}).
\begin{equation}
\begin{aligned}
&\ln Z_s^{(1)}=\ln N!+N\ln(4\pi)+\\
&\sum_{i=1}^N \{\ln[m(i-1)]!+\ln[m(N-i)]!-\ln[m(N-1)+1]!\}
\label{eq: appendix_Zs_scaling}
\end{aligned}
\end{equation}
As described in the disk case, the factorials can be approximated using the Stirling's formula
\begin{equation}
\ln N!=N\ln N-N+\frac{1}{2} \ln N+\frac{1}{2} \ln (2\pi)+\frac{1}{12N}+\cdots.
\end{equation}
And the summation could be converted to integral using the Euler-Maclaurin formula Eq.(\ref{eq: appendix_Euler_Maclaurin}) where $f(i)$ is now defined as follows,
\begin{widetext}
\begin{equation}
\begin{aligned}
&f(i)=\ln[m(i-1)]!+\ln[m(N-i)]!\\
=&\left\{m(i-1)\ln [m(i-1)]-m(i-1)+\frac{1}{2} \ln [m(i-1)]+\frac{1}{2}\ln(2\pi)+\frac{1}{12 m(i-1)}\right\}+\\
&\left\{m(N-i)\ln [m(N-i)]-m(N-i)+\frac{1}{2} \ln [m(N-i)]+\frac{1}{2}\ln(2\pi)+\frac{1}{12 m(N-i)}\right\}.
\end{aligned}
\end{equation}
\end{widetext}
And its derivative is
\begin{equation}
\frac{df(i)}{di}=m\ln[m(i-1)]+\frac{1}{2(i-1)}-m\ln[m(N-i)]+\cdots.
\end{equation}
Plug these into the Euler-Maclaurin formula, we find the summation part of Eq. (\ref{eq: appendix_Zs_scaling})
\begin{widetext}
\begin{equation}
\begin{aligned}
&\sum_{i=1}^N \{\ln[m(i-1)]!+\ln[m(N-i)]!-\ln[m(N-1)+1]!\}\\
=&\sum_{i=1}^Nf(i)-N\{[m(N-1)+1]\ln [m(N-1)+1]-[m(N-1)+1]+\\
&\frac{1}{2} \ln [m(N-1)+1]+\frac{1}{2}\ln(2\pi)+\frac{1}{12 [m(N-1)+1]}\}+O(1)\\
=&-\frac{m}{2}N^2-\frac{1}{2} N\ln N+\frac{1}{2}[2(m-1)+\ln(2\pi)]N+\frac{m^2-3m+1}{6m}\ln N+O(1).
\end{aligned}
\end{equation}
\end{widetext}
Inserting this result in the expression of $B(N)Z_s^{(1)}$ one can easily find the overlap scaling as Eq. (\ref{eq: sphere_scaling}) in the main text.

\providecommand{\noopsort}[1]{}\providecommand{\singleletter}[1]{#1}%

\end{document}